\documentclass[preprint2]{aastex}

\usepackage{aastexug}

\begin{document}

\title{Dust Temperatures in the Infrared Space Observatory\footnote{Based
on observations with ISO, an ESA project with instruments funded by
ESA Member States (especially the PI countries: France, Germany, the
Netherlands and the
United Kingdom) and with the participation of ISAS and NASA.} Atlas of Bright
Spiral Galaxies}

\author{George J. Bendo,\altaffilmark{2,3,4}
Robert D. Joseph,\altaffilmark{3}
Martyn Wells,\altaffilmark{5} Pascal Gallais,\altaffilmark{6}
Martin Haas,\altaffilmark{7} Ana M. Heras,\altaffilmark{8, 9}
Ulrich Klaas,\altaffilmark{7} Ren\'{e} J. Laureijs,\altaffilmark{8, 9}
Kieron Leech,\altaffilmark{9, 10} Dietrich Lemke,\altaffilmark{7}
Leo Metcalfe,\altaffilmark{8}
Michael Rowan-Robinson,\altaffilmark{11} Bernhard Schulz,\altaffilmark{9, 12}
and Charles Telesco\altaffilmark{13}}

\altaffiltext{2}{Steward Observatory, 933 North Cherry Ave, Tucson, AZ 85721
USA; gbendo@as.arizona.edu}
\altaffiltext{3}{University of Hawaii, Institute for Astronomy,
2680 Woodlawn Drive, Honolulu, HI 96822, USA; joseph@ifa.hawaii.edu}
\altaffiltext{4}{Guest User, Canadian Astronomy Data Center, which is
operated by the Dominion Astrophysical Observatory for the National Resarch
Council of Canada's Herzberg Institute of Astrophysics. }
\altaffiltext{5}{UK Astronomy Technology Center, Royal Observatory Edinburgh,
Blackford Hill, Edinburgh EH9 3HJ, Scotland, UK; mw@roe.ac.uk}
\altaffiltext{6}{CEA/DSM/DAPNIA Service d'Astrophysique,
F-91191 Gif-sur-Yvette, France; gallais@discovery.saclay.cea.fr}
\altaffiltext{7}{Max-Planck-Institut f\"{u}r Astronomie,
K\"{o}nigstuhl 17, D-69117, Heidelberg, Germany;
haas@mpia-hd.mpg.de, klaas@mpia-hd.mpg.de, lemke@mpia-hd.mpg.de}
\altaffiltext{8}{Astrophysics Division, Space Science Department of ESA,
ESTEC, P.O. Box 299, 2200 AG Noordwijk, Netherlands;
aheras@estsa2.estec.esa.nl, rlaureij@rssd.esa.int}
\altaffiltext{9}{ISO Data Center, Astrophysics Division, ESA,
Villafranca del Castillo, 28080 Madrid, Spain;
lmetcalf@iso.vilspa.esa.es}
\altaffiltext{10}{Said Business School, Park End Street, Oxford OX1 1HP,
England, UK; kieron.leech@said-business-school.oxford.ac.uk}
\altaffiltext{11}{Astrophysics Group, Imperial College, Blackett Laboratory,
Prince Consort Road, London SW7 2BZ, England, UK;
m.rrobinson@ic.ac.uk}
\altaffiltext{12}{IPAC, California Institute of Technology 100-22, Pasadena,
CA 91125 USA; bschulz@ipac.caltech.edu}
\altaffiltext{13}{Department of Astronomy, University of Florida,
P.O. Box 112055, Gainesville, Florida 32611, USA;
telesco@astro.ufl.edu}

\shorttitle{Dust Temperatures in ISO Atlas}
\shortauthors{Bendo et al.}

\begin{abstract}
We examine far-infrared and submillimeter spectral energy distributions for
galaxies in the Infrared Space Observatory Atlas of Bright
Spiral Galaxies.  For the 71 galaxies where we had complete 60~- 180~$\mu$m
data, we fit blackbodies with $\lambda^{-1}$ emissivities and average
temperatures of 31~K or $\lambda^{-2}$ emissivities and average temperatures
of 22~K.  Except for high temperatures determined in some early-type
galaxies, the temperatures show no dependence on any galaxy characteristic.
For the 60 - 850~$\mu$m range in eight galaxies, we fit blackbodies with
$\lambda^{-1}$, $\lambda^{-2}$, and $\lambda^{-\beta}$ (with $\beta$ variable)
emissivities to the data.  The best results were with the $\lambda^{-\beta}$
emissivities, where the temperatures were $\sim$~30~K and the
emissivity coefficient $\beta$ ranged from 0.9 to 1.9.  These results
produced gas to dust ratios that ranged from 150 to 580, which were
consistent with the ratio for the Milky Way and which exhibited
relatively little dispersion compared to fits with fixed emissivities.
\end{abstract}


\section{Introduction}

The far-infrared emission from galaxies is emitted by interstellar dust
heated by a variety of energy sources.  Two important energy sources are
the general interstellar radiation field and the more intense radiation in
star formation regions.  These should produce dust sources at
different dust temperatures and with different spatial distributions.
Understanding these  dust temperature components is necessary for
understanding the relation between far-infrared emission and star formation
in galaxies.

While much of the research in this subject has focused on ultraluminous
galaxies and high redshift objects,  some studies have investigated dust
temperatures in normal spiral galaxies.  However, no clear consensus has
been reached on the temperatures of the dust components in spiral
galaxies.  Perhaps the most contentious issue is whether a significant dust
component at a very low temperature $\sim$ 10-15~K exists. This issue needs
to be resolved to understand how the interstellar  radiation field and star
formation regions heat dust to produce far-infrared emission.

Before the launch of the Infrared Space Observatory (ISO) \citep{ketal96},
most research on dust temperatures in galaxies relied on combining IRAS or
Kuiper Airborne Observatory (KAO) data  with ground-based submillimeter or
millimeter data.  The launch of ISO has permitted better far-infrared
spectral energy distributions to be measured.

\citet{s82} was the first to measure a dust temperature in a non-active
spiral.  Using KAO data, he found a temperature of 20 K in NGC 5194
(M 51).  \citet{cetal86} combined submillimeter bolometer data from the NASA
IRTF with IRAS far-infrared data to conclude that spiral galaxies have
cold dust components with an average temperature of 16 K.  \citet{setal89}
measured far-infrared fluxes with the KAO and submillimeter fluxes with the
NASA IRTF for three Virgo cluster spirals and inferred significantly warmer
dust temperatures than those determined by \citet{cetal86}.  \citet{ewd89}
measured dust temperatures in 11 galaxies using IRAS far-infrared data and
submillimeter bolometer data from UKIRT and determined that the coolest dust
component in spiral galaxies was 30 - 50 K, much warmer than previous
measurements.  These studies thus established a controversy over the
apparently simple issue of the dust temperatures in spiral galaxies.

Subsequent studies have produced similarly conflicting results.  Most have
focused on individual galaxies, including NGC 660, UGC 3490
\citep{ck93}, NGC 3267 \citep{srhetal94}, NGC 4631 \citep{betal95}, M 51
\citep{getal95, hetal96}, M~101 \citep{hetal96}, the Andromeda
Galaxy \citep{hlsetal98}, and NGC~891 \citep{petal00}.  These studies
typically find dust temperatures of $\sim$~20 K, with the lowest temperature
being 16~K and the highest being 33~K.  There have been a few studies of larger
samples. Some of these surveys have found far-infrared dust temperatures as
high as 35~K \citep{cac93, detal00}. Others have found temperatures between
20 and 30~K \citet{de01, petal02}, and still others have found dust
components as cold as 10-15~K \citep{cetal95, ketal98, skc99, setal00}.

Part of the reason no consensus has been reached on dust temperatures in
normal galaxies is because of poor sampling of the spectral energy
distribution (SED). To find reliable dust temperatures it is vital to evenly
sample the SEDs from the Wien side to the Rayleigh-Jeans side of the Planck
spectrum.  Some studies have had such poor spectral sampling that Planck
functions are fit to data with no degrees of freedom, and some have entire
dust temperature components defined by a single measurement. Another source
of the discrepancy is that  different studies assume different wavelength
dependences for the emissivity.  In general when a shallower emissivity
function is chosen, this produces warmer temperatures.

In this paper we combine far-infrared ISO photometry with submillimeter
photometry from the Submillimetre Common-User Bolometer Array (SCUBA)
\citep{hrgetal99} at the James Clerk Maxwell Telescope (JCMT) to sample the
full spectrum of cool dust emission in normal spiral galaxies.  We will
first present the dust temperatures we infer from 60, 100, and 180~$\mu$m
ISO measurements  for 71 spiral galaxies in the ISO Atlas of Bright Spiral
Galaxies \citep{betal02} (henceforth referred to as Paper 1).  We then
examine dust temperatures obtained by combining the far-infrared data with
submillimeter data for 8 of the galaxies in the sample. Finally, we
calculate dust masses and examine gas to dust mass ratios for 7  of
these galaxies.

\section{Data}

\subsection{The Sample}

The galaxies in this sample are a subset of a complete, magnitude-limited set
of galaxies selected from the Revised Shapley-Ames (RSA) Catalog \citep{st87}.
The sample comprised galaxies with Hubble types between S0 and Sd and with
magnitudes B$_T$~=~12 or brighter; galaxies in the Virgo Cluster were
excluded.  A randomly selected, subset of these galaxies was observed by ISO
based on target visibility.  This produced a total of 77 galaxies that are
representative of the range of Hubble types in the RSA Catalog, 71 of
which were observed successfully with ISOPHOT.  Detailed information on the
sample is presented in Paper 1.

\subsection{Observations and Data Analysis}

\subsubsection{Far-Infrared Photometry}

The 60, 100, and 180~$\mu$m photometric data were taken with ISOPHOT
\citep{letal96} and processed with PIA 8.0 \citep{getal97} with additional
processing to correct for PSF effects, as discussed in Paper 1. Note that,
for the analysis without the submillimeter data, we assume that  the
180~$\mu$m flux within the 180~arcsec apertures effectively comes from the
same sources as the 60 and 100~$\mu$m fluxes within a 135~arcsec aperture.
To investigate how much extra far-infrared flux may  come from the region
outside the 135~arcsec aperture but within the 180~arcsec aperture, we
investigated the distribution of infrared flux in maps of IRAS data made by
the HIRES software package provided by the Infrared Processing and Analysis
Center.  We find that the median ratio of flux within 135~arcsec to flux
within 180~arcsec is  0.85 at 60~$\mu$m and 0.80 at 100~$\mu$m for the
galaxies in this sample.   According to the HIRES beam profiles, point
sources would have ratios of 0.95 at 60~$\mu$m and 0.85 at 100~$\mu$m.
The small difference between the ratio of fluxes for the sample galaxies and
the ratio expected for point sources suggests that most of the emission
between the 135 and 180~arcsec apertures in the HIRES maps is mostly from
the point spread  function and that emission in those regions is relatively
small in most  galaxies.  We will assume that the distribution of the
180~$\mu$m emission will be similar to the 100~$\mu$m emission, so
therefore, we will assert that most of the 180~$\mu$m emission for the
galaxies in this sample also comes from within 135~arcsec. When we also
have submillimeter data, however, we can make assumptions about the spatial
distribution of the emission and  perform a deconvolution analysis to
estimate the 180~$\mu$m fluxes within 45 and 135~arcsec apertures.

\subsubsection{Submillimeter Photometry}

We observed NGC~4088 and NGC~4826 at 850 $\mu$m using SCUBA at the JCMT.
The data were taken in jiggle map mode, in which the telescope's secondary
mirror shifts in a 64-point pattern to create a completely sampled map for the
inner 135 arcsec of the target.   For sky subtraction, we nodded 2~arcmin
off-source parallel to the optical disks' minor axes for each galaxy.
Integration times were 64 min for NGC~4088 and 60 min for NGC~4826.
We performed skydips regularly to check the opacity during the
observations.  At 850 $\mu$m, the opacity  typically ranged between
0.3 and 0.6 for all the observations.  (The high sky opacity made the
simultaneous 450 $\mu$m observations useless.)  For calibration, we observed 
Mars using similar methods.

In addition to the above data, we have used data from the JCMT SCUBA
archives for 6  galaxies: NGC 4414, NGC 4631, NGC 5236, NGC 5713, NGC~5792,
and NGC~5907.  All of these data were taken in SCUBA's jiggle map
mode, with either Uranus, OH~231.8, or IRC~+10216 used for calibration.

The data were processed with the SCUBA User Reduction Facility
\citep{jl98} following the typical processing steps. First, the 
data were flat-fielded and corrected for airmass and sky opacity.
Bad  bolometers were flagged and masked.   An additional background selection
was made by measuring and subtracting surface brightnesses in selected
off-target bolometers using the "remsky" procedure.   Finally, the data were
combined together and rbinned to produce final maps. Images of planets and 
other calibration sources were processed using the same methods.  Signals were
measured within 45 and 135 arcsec apertures, which are comparable to the
apertures of ISOPHOT.  For calibration, we measured the signals from the
standard sources in similarly-sized apertures, and these then provided
conversion factors for converting signals to flux densities for the targets.
We assume that the uncertainty in  both the new measurements and the
archival measurements comes primarily from  the uncertainty in the
calibration, which is approximately 10~\%.  Other  sources of uncertainty,
mainly uncertainties from background noise, appear to  be less significant.
Table~\ref{t_smm} lists the resulting flux densities as  well as the optical
diameter parameter from the Third Reference Catalogue of  Bright Galaxies
\citep{detal91} and an indication of whether the data came from original
observations or the JCMT SCUBA archives.

\section{Spectral Energy Distributions and Dust Temperatures}

\subsection{Far-Infrared Spectral Energy Distributions}

First, we fit blackbody functions with two different emissivity functions to
the data. The  emissivity functions we selected scale as $\lambda^{-1}$
and $\lambda^{-2}$, functions that are used by many studies cited in
Section 1 and that are recommended by \citet{h83}.

\begin{figure}
\plottwo{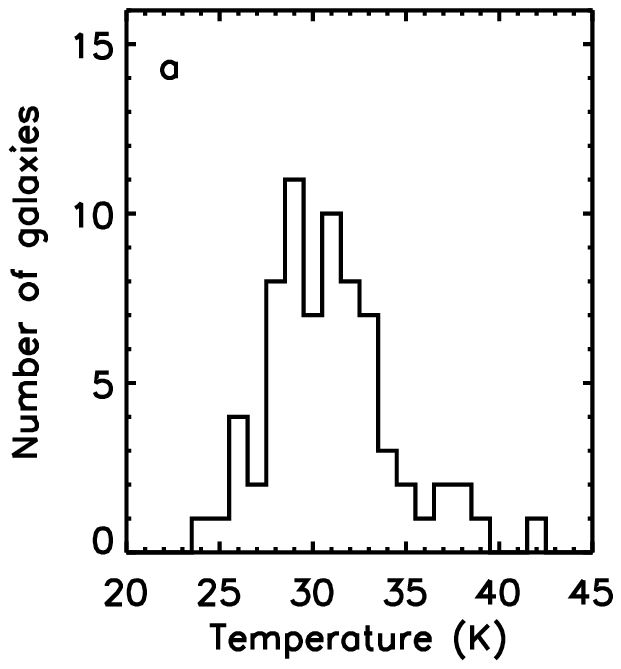}{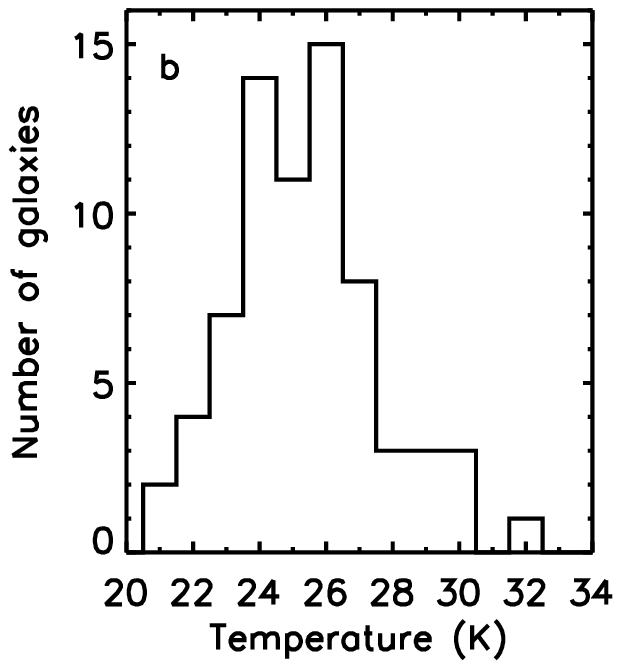}
\caption{Histograms of the temperatures with (a) $\lambda^{-1}$ and
(b) $\lambda^{-2}$ emissivities, as determined from the 60 - 180~$\mu$m data.}
\label{f_temphist}
\end{figure}

Table~\ref{t_firsed} lists the dust temperatures determined from the ISO
far-infrared data, and Figure~\ref{f_temphist} shows histograms of these
temperatures.  We estimate an error of $\sim$~2~K in the dust
temperatures due to uncertainties in the photometric calibration of
ISOPHOT.  This error was determined by fitting blackbodies to data where the
60 $\mu$m flux was increased 20~\% and the 180 $\mu$m flux was decreased 20
\% and vice-versa. The temperatures of the fits are, for the the
$\lambda^{-1}$ emissitivity, a mean of 31~K with a range of 24~- 42~K, and,
for the $\lambda^{-2}$ emissitivity, a mean of 25~K with a range of 21~-
32~K.  The functions with the $\lambda^{-1}$ emissitivities generally fit
the data better according to the $\chi^2$ criterion.

The dust temperatures vary little in relation to morphology and AGN activity.
The data in Table~\ref{t_firsedmorph} do show that some S0 and elliptical
galaxies tend to have warmer dust temperatures, but all the spiral and
irregular galaxies have indistinguishable dust temperatures.

 These results should be compared with those of
\citet{setal00}, who used IRAS 60 and 100~$\mu$m data supplemented by 170
$\mu$m photometry from the ISO Serendipity Survey for 115 galaxies. They
assumed an emissivity function scaling as $\lambda^{-2}$ and found a
temperature distribution centered at 20~K with a range of 15-25K.  These
results are systematically colder by 5-6~K and they conclude that "for the
first time this indicates that a cold dust component with T$_D \le$~20~K is
present in a large fraction of all spiral galaxies."  In contrast, using the
same emissivity function we find no galaxy with T$_D <$~21~K and no support
for such a cold dust component in most spiral galaxies.  One possible
explanation could be that \citet{setal00} use IRAS Faint Source Catalog (FSC) 
data at 60 and
100~$\mu$m, whereas we have used ISO photometry at 60 and 100~$\mu$m and were
able to make approximate corrections for the point spread function.  As we 
demonstrated in Paper 1, the ISO fluxes used in this research, particularly the
60~$\mu$m fluxes, are systematically higher than those in the IRAS FSC 
Catalog, which would naturally lead to the temperatures found using the ISO 
photometry to be higher than those found using corresponding IRAS FSC data.

However, some qualifications must be made to interpretations
based on data confined to this relatively small wavelength range, even
though these data from ISO include the longer wavelength 180~$\mu$m
measurement which IRAS did not have. First, if dust components 10-15~K
exist, their contribution would not be evident using the far-infrared data
alone.  Second, emission at 60~$\mu$m may include emission from warmer dust 
components that cause the spectral energy distribution to flatten out 
(see \citet{letal00} for an  example
based on observational data of galactic cirrus, \citet{petal00} and
\citet{de01} for examples of blackbody fits to observations of nearby
galaxies, or \citet{dbp90} and \citet{detal01} for detailed modeling of dust 
emission at far-infrared wavelengths). Thus, fitting single blackbody 
functions through the 60, 100, and 180~$\mu$m data points is possible but 
may not accurately describe the physics of the dust emission.
The addition of longer wavelength data may provide evidence for more than one
dust component.  Therefore, in the next section, we examine the blackbody
fits obtained when the  far-infrared data are combined with submillimeter
data.

\subsection{Far-Infrared to Submillimeter Spectral Energy Distributions}

We have submillimeter and far-infrared data for a subset of 8 of the ISO
galaxy sample, and blackbody functions with $\lambda^{-1}$ and
$\lambda^{-\beta}$ emissivities, with $\beta$ variable, have been fitted to
these data. We refer to the resulting dust temperatures measured within
135~arcsec apertures as global dust temperatures.  We also
determined the temperatures of dust within the inner 45~arcsec of
the galaxies, and we refer to these as the nuclear dust temperatures. (We
note our designation "nuclear" is very qualitative; at the distances of
these galaxies 45 arcsec subtends $\ge$ 1 kpc, a much larger region than
would normally be considered the galaxy nucleus. Starburst regions in
galaxies are typically 250 - 500 pc.)   The dust temperatures measured within
a 135~arcsec aperture but outside a 45~arcsec aperture we refer to as disk
temperatures.  These distinctions permit investigation of the differences
between dust temperatures in the inner and outer regions of these galaxies.

Since the diffraction limit of the ISO telescope at 180~$\mu$m is $\sim$~145
arcsec, to do this analysis for the 180~$\mu$m data requires estimating the
fluxes from both central unresolved and extended components.  
We use a method similar to that used by \citet{retal99}. (We also
used this method for the 60 and 100~$\mu$m data in Paper 1.)  We calculate
the ratio of fluxes for the point and extended sources at both 100~$\mu$m and
850 $\mu$m, since we have the required 45 arcsec resolution at both these
wavelengths.  We then interpolate to estimate the ratio of the 180~$\mu$m
fluxes from point and extended components.  According to
\citet{l99}\footnote{http://www.iso.vilspa.esa.es/users/expl\_lib/PHT/c200fpsf02.ps.gz}, 

61.2~\% of the flux from a point source centered in the C200 array (the 180
$\mu$m measurement) will fall on the array.  Therefore the flux within the
180~arcsec aperture can be expressed as
\begin{equation}
f_m(180\arcsec) = 0.612 f_c + (180 \arcsec)^2 f_e .
\end{equation}
where f$_m$ is the measured flux, f$_c$ is the flux from the central source,
and f$_e$ is the flux per arcsec$^2$ from a diffuse, uniform, extended
source.  Substituting the ratio $\frac{f_c}{(180 \arcsec)^2 f_e}$ into the
above equation allows for solving for f$_c$ and f$_e$:
\begin{equation}
f_c = \frac{f_m}{0.612+\frac{(180 \arcsec)^2 f_e}{f_c}}
\end{equation}
\begin{equation}
f_e = \frac{f_m}{0.612\frac{f_c}{(180 \arcsec)^2 f_e}+1}
\end{equation}
These central and extended components can then be used to calculate fluxes
within 45 and 135~arcsec apertures:
\begin{equation}
f_{t}(45\arcsec) = f_c + (45\arcsec)^2 f_e
\end{equation}
\begin{equation}
f_{t}(135\arcsec) = f_c + (135\arcsec)^2 f_e.
\end{equation}
The resulting 180~$\mu$m flux desnities are listed in Table~\ref{t_180}.
We can now use these 180~$\mu$m flux estimates for the dust temperature
analysis outlined above.

\begin{figure}
\plotone{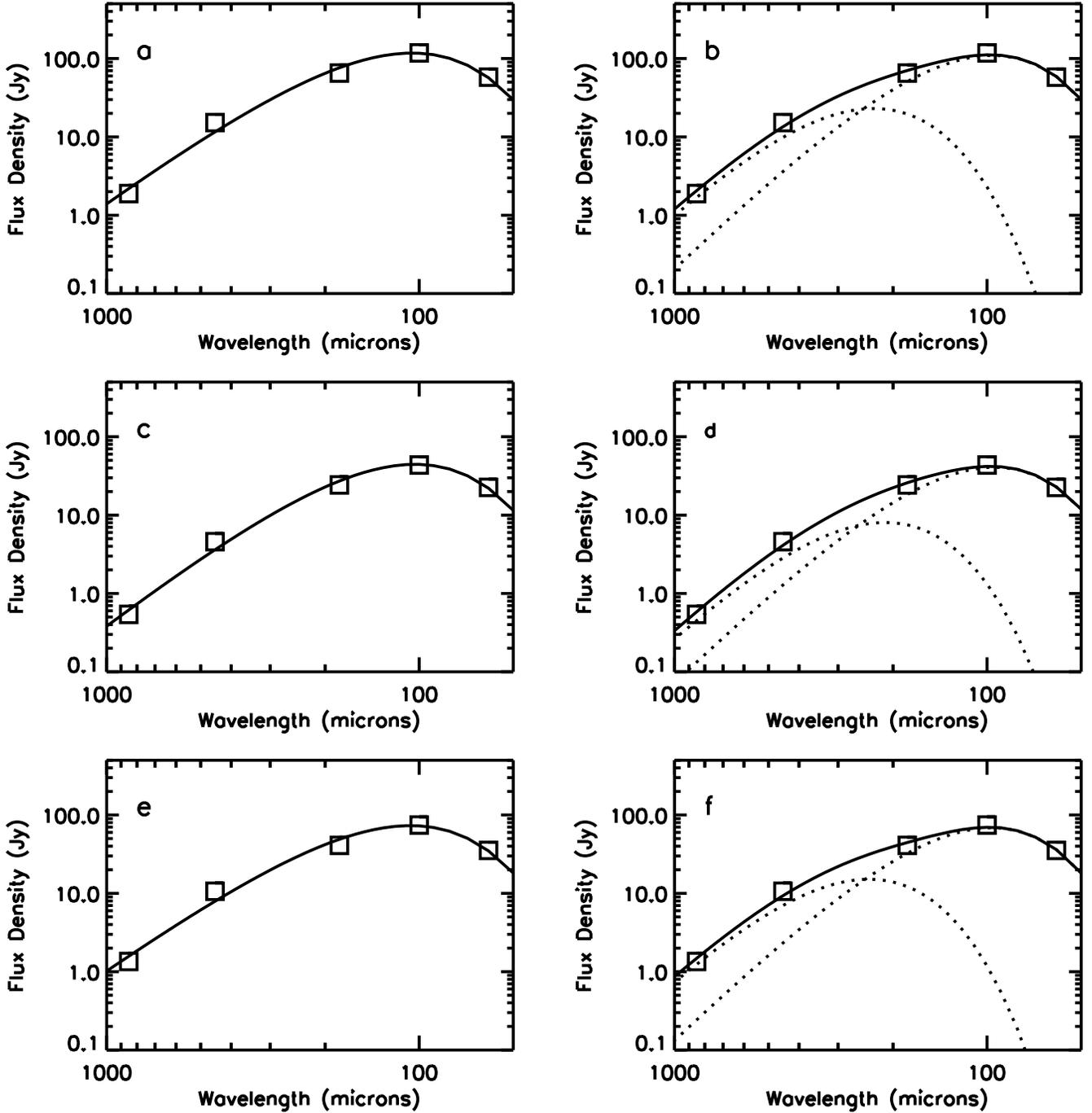}
\caption{The (a \& b) global, (c \& d) nuclear, and (e \& f) disk spectral
energy distributions for NGC~4631.  The measured data are represented by
squares, and the errors for the data are less than the size of the squares.
Parts a, c, and e show the best fitting blackbodies with
$\lambda^{-\beta}$ emissivities (with $\beta$ variable).  Parts b, d, and f
show the best fitting two blackbody components with
$\lambda^{-2}$ emissivities, with the dotted lines represent the individual
blackbody components.  This format also applies to Figures~\ref{f_sed4088}
- \ref{f_sed5907},
although only the global spectral energy distributions are shown in those
cases.}
\label{f_sed4631}
\end{figure}

\begin{figure}
\plottwo{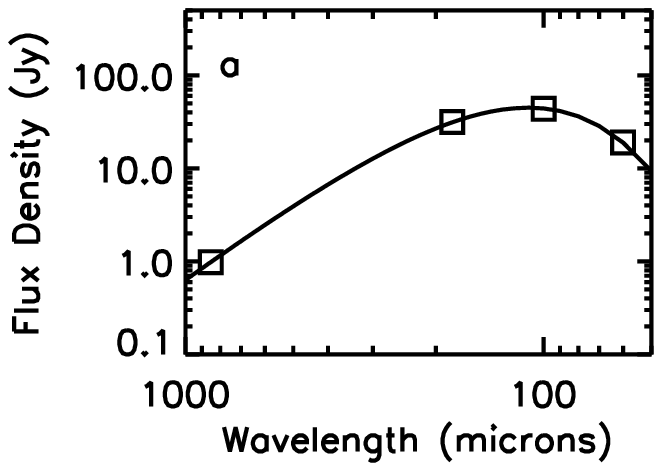}{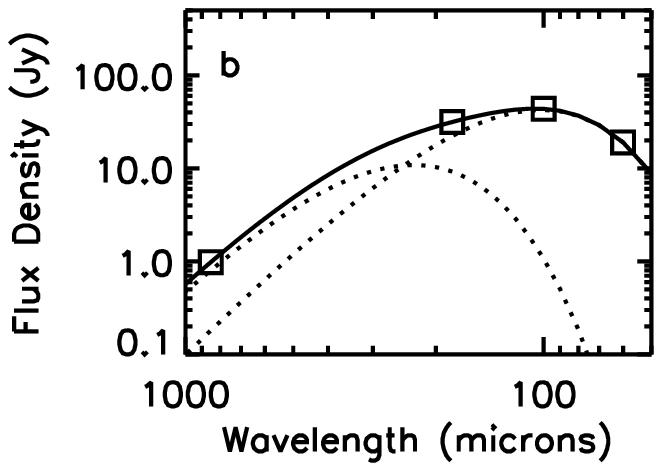}
\caption{The global spectral energy distribution for NGC~4088.}
\label{f_sed4088}
\end{figure}

\begin{figure}
\plottwo{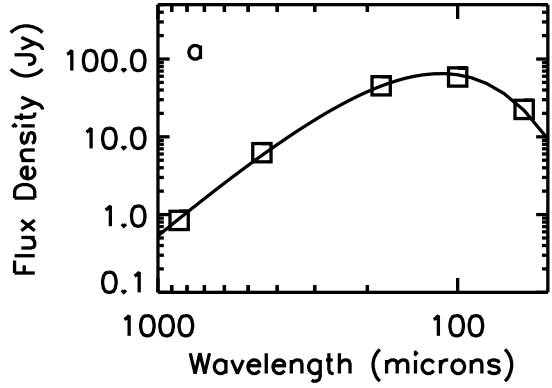}{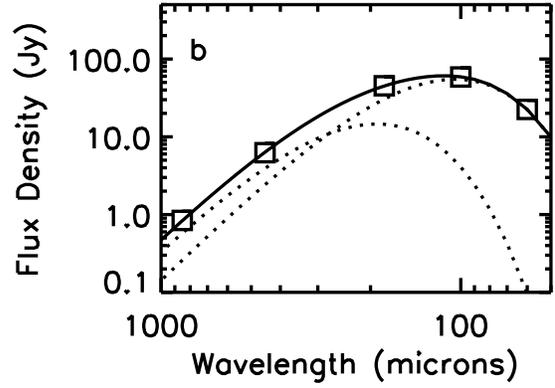}
\caption{The global spectral energy distribution for NGC~4414.}
\label{f_sed4414}
\end{figure}

\begin{figure}
\plottwo{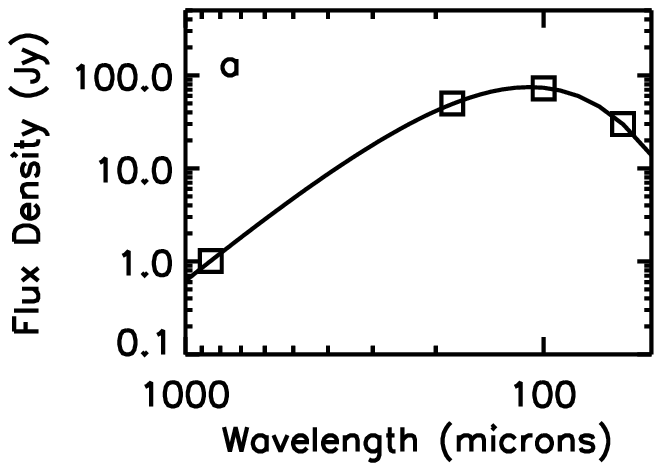}{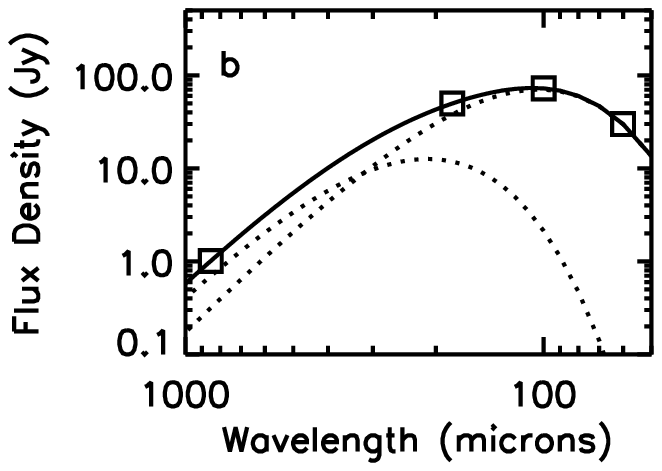}
\caption{The global spectral energy distribution for NGC~4826.}
\label{f_sed4826}
\end{figure}

\begin{figure}
\plottwo{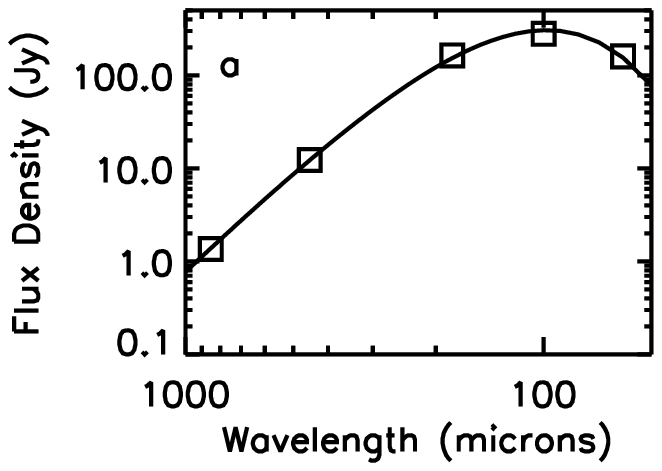}{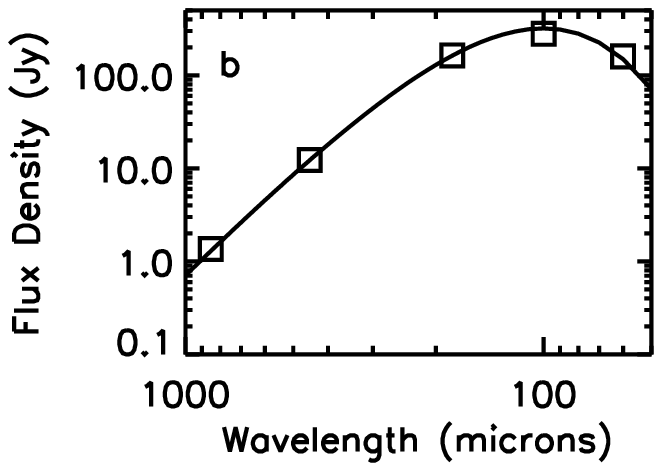}
\caption{The global spectral energy distribution for NGC~5236.}
\label{f_sed5236}
\end{figure}

\begin{figure}
\plottwo{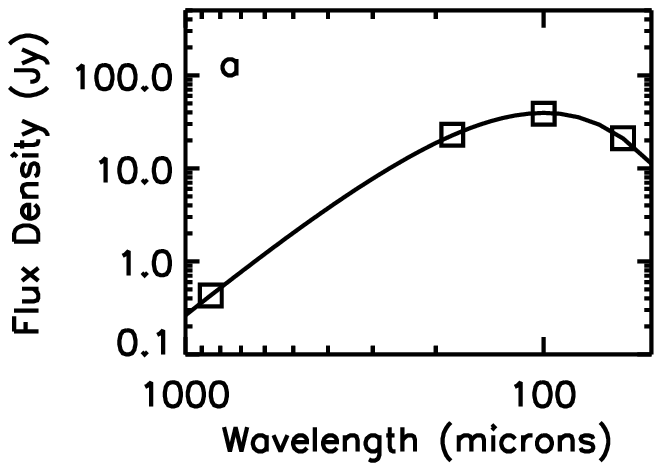}{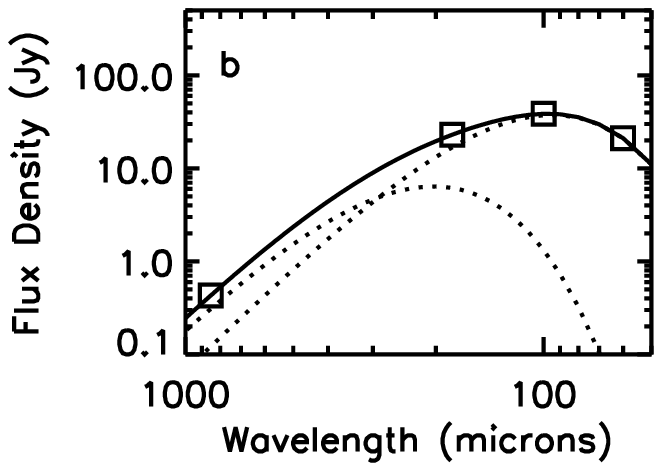}
\caption{The global spectral energy distribution for NGC~5713.}
\label{f_sed5713}
\end{figure}

\begin{figure}
\plottwo{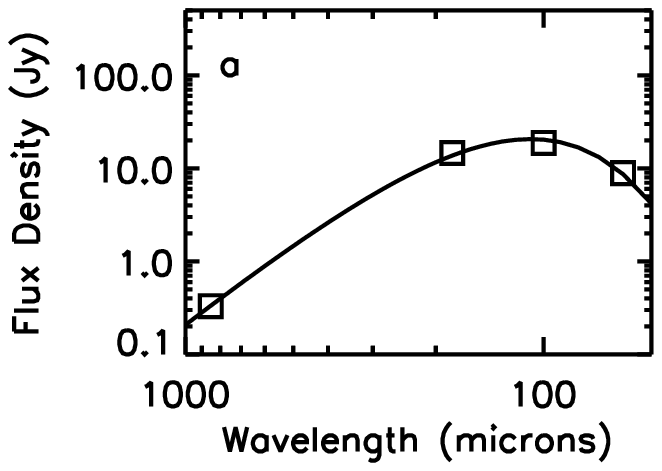}{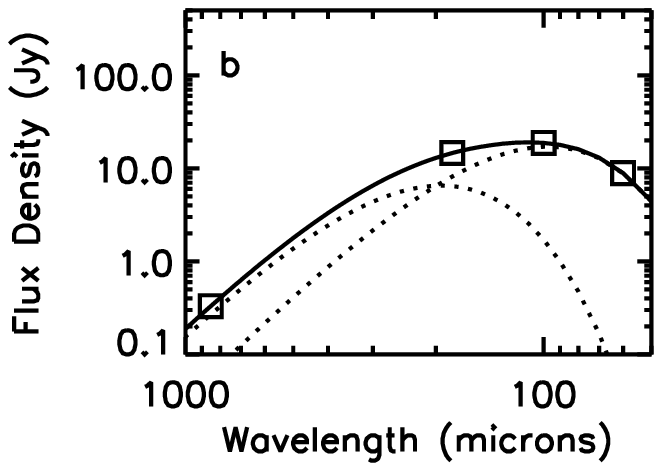}
\caption{The global spectral energy distribution for NGC~5792.}
\label{f_sed5792}
\end{figure}

\begin{figure}
\plottwo{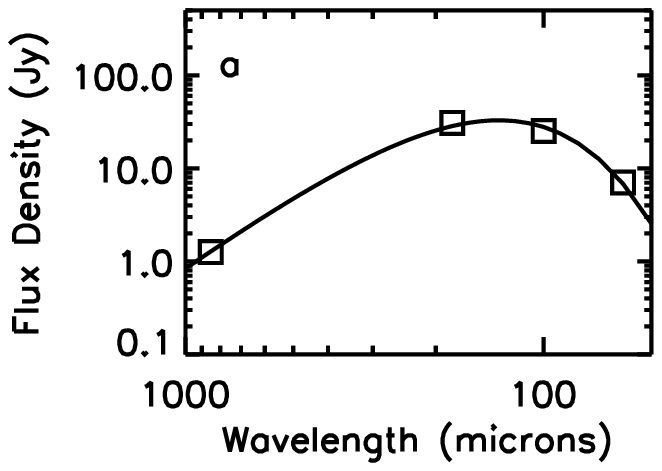}{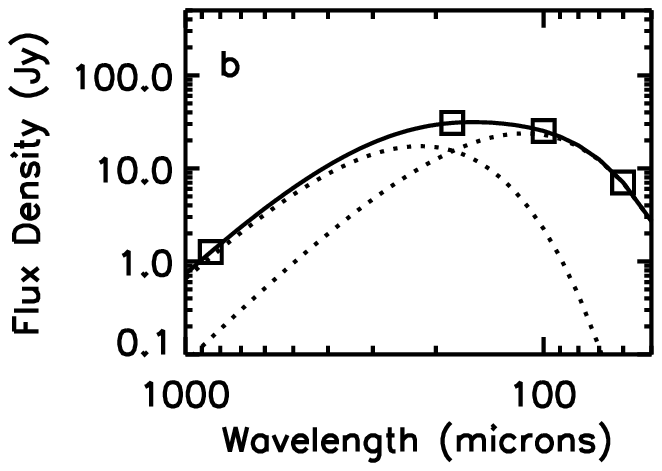}
\caption{The global spectral energy distribution for NGC~5907.}
\label{f_sed5907}
\end{figure}

Table~\ref{t_firsmmsed_1} lists the resulting dust temperatures derived using
the $\lambda^{-1}$ emissivities, and Table~\ref{t_firsmmsed_n} lists the dust
temperatures for $\lambda^{-\beta}$ emissivities.
In addition, Figures~\ref{f_sed4631} - \ref{f_sed5907} show the spectral
energy distributions and the fits of blackbody functions with
$\lambda^{-\beta}$ emissivities.  We also tried fitting single blackbodies
with $\lambda^{-2}$ emissivities to the data, but except for NGC~5236 (which
will be discussed below) the fits were very poor.   We estimated an error of
$\sim$~1~K in these dust temperatures using the same method described above
for the far-infrared spectral energy distributions.  In summary, the fits
yield dust temperatures of $\sim$~35~K for models with $\lambda^{-1}$
emissivities and $\sim$~30~K for models with $\lambda^{-\beta}$
emissivities.  Typical fits to the emissivity coefficient were between 0.9
and 1.4.

NGC~5236 is exceptional in that the emissivity function is closer to
$\lambda^{-2}$.  NGC~5236 is a galaxy with unusual supernova activity.  Shocks
from the strong supernova activity associated with the galaxy's nucleus could
be responsible for breaking up larger dust grains into smaller grains, thus
making the wavelength dependence of the emissivity function steeper.  The
disk of NGC~4414, but not the nucleus, also exhibits a steep emissivity.
Star formation has been inhibited in the nucleus by the lack of molecular
gas present \citep{s96}, but sites of strong star formation are evident in
the disk ({\it cf.} Fig. 25 in Paper 1), and the large dust grains may have
suffered similar fragmentation as in NGC~5236.

In general, the "nuclear" dust temperatures are systematically warmer
than the disk temperatures in all cases.  Because stellar densities are
higher in the nuclei of these galaxies, and because the star formation rate
may be higher, the interstellar radiation field is stronger.  This enhances
both the dust emission and the dust temperatures, producing the observed
differences in dust temperatures between the nuclear and disk regions.

Since some other studies have suggested there are two temperature components,
we have also investigated the possibility that the dust emission may be
composed of two blackbody components.  This obviously requires steeper
wavelength-dependence for the emissivities.  We therefore chose emissivity
functions proportional to $\lambda^{-2}$ for both temperature components.
Table~\ref{t_firsmmsed_2} shows the best fitting temperatures, with two
temperatures derived for all galaxies except NGC~5236.  (For
NGC~5236, the analysis demonstrates that a second blackbody component would
make only a minor  contribution to the far-infrared and submillimeter
spectral energy  distribution, so only one component was fit to the data.)
In most cases, the two components have temperatures of $\sim$~15~K and
$\sim$~30~K.  For galaxies with no 450~$\mu$m data, the solution is an
even-determined problem with an exact fit that should be interpreted with
caution.  Moreover, the colder dust temperature is dependent mainly on the
submillimeter measurement, so errors in the submillimeter measurement
directly affect the temperature of the coldest component.

Having used submillimeter data to determine dust temperatures for these 8
galaxies, we can now compare the temperatures found for these same galaxies
used the ISO far-infrared data alone. Comparing the dust temperatures in
Table~\ref{t_firsed} with the "Global" temperatures in
Table~\ref{t_firsmmsed_1} and Table~\ref{t_firsmmsed_2} shows that the
addition of the submillimeter data systematically increases the derived dust
temperatures $\sim$~2-3~K.  The more significant difference is that, by
adding the submillimeter data point(s) one can search for a very cold dust
component of temperature $\sim$~6-15 K.

In summary, for these 8 galaxies with both ISO far-infrared and submillimeter
measurements, we find dust temperatures in almost all cases that can be
described as a $\sim$~35~K blackbody with a $\lambda^{-1}$ emissivity, a
$\sim$~30~K blackbody with a variable emissivity coefficient ranging from
0.9 to 1.4, or two blackbodies with temperatures of $\sim$~15~K and
$\sim$~30~K and $\lambda^{-2}$ emissivities.  Clearly the
far-infrared-to-submillimeter spectral energy distributions of galaxies can
be represented with several different functions and correspondingly larger or
smaller dust temperatures.  This is essentially a duplication of the variety
of results found in the literature for the past two decades. The key
question is, which of these is most likely to be correct?  We return to this
question in the Discussion.

\section{Gas to Dust Mass Ratios}

The accurate measurement of dust temperatures is not only necessary for
understanding dust emission in galaxies.  It is also required for
measuring the gas to dust ratios in spiral galaxies.  As discussed by
\citet{dy90}, the temperatures and dust masses measured with IRAS data, when
combined with molecular gas data, produced gas to dust mass ratios of
approximately 1000.  This is a factor of 10 larger than the value for the
Milky Way and it therefore raises a serious problem. This problem has also
appeared in most other studies in which cold dust components have been
found.  The ISO Serendipidy Survey by \citet{setal00} is a notable
exception; the typical gas-to-dust mass ratios found are near that of the
Milky Way, but the ratios range from 50 to 1000. If these results are
reliable they raise the question whether this basic property of the
interstellar medium really varies so dramatically among galaxies.

\citet{yetal95} list CO flux measurements for the 8 galaxies for which we
have far-infrared and submillimeter flux measurements. Using their data and
their conversion of CO intensities to molecular gas surface densities, we
calculated molecular gas densities.  We then used the 850~$\mu$m flux
densities and the dust temperatures and the associated emissivity
assumptions used to the fit the SEDs to calculate dust masses. Working
at 850~$\mu$m is preferred because the determination of dust masses at that
wavelength is relatively insensitive to uncertainties in dust temperatures.
With these masses, we then calculated gas to dust ratios that may be
compared to the Milky Way ratio.

\subsection{Dust Masses and Gas to Dust Mass Ratios}

We calculated dust masses using
\begin{equation}
M_{Dust} = [\frac{F(\nu)D^2}{B(\nu,T)}][\frac{4a}{3Q(\nu)}]\rho
\end{equation}
derived in \citet{h83}.
M$_d$ is the dust mass, F($\nu$) is the measured flux density, D is
the distance, B($\nu$,T) is the flux density from a blackbody of temperature
T, a is the approximate average size of an interstellar dust grain,
Q($\nu$) is the emissivity, and $\rho$ is the density of a dust grain.
Since these are nearby spiral galaxies, we assume that these galaxies
contain dust with properties similar to the dust in the Milky Way.  Therefore,
we have also used the asssumptions in \citet{h83} for calculating dust masses.
The size and density of the dust grains are assumed to be a~=~0.1~$\mu$m and
$\rho$~=~3~g~cm$^{-3}$.
For the emissivities, we use the expression
\begin{equation}
Q(\lambda,\beta) = Q_0a(\frac{250}{\lambda})^\beta
\end{equation}
from \citet{ketal01}, where Q$_0$~=~40~cm$^{-1}$ and $\beta$ is the emissivity
coefficient.  This equation gives
Q(850$\mu$m,1) = 1.2 $\times$~10$^{-4}$ and
Q(850$\mu$m,2) = 3.5 $\times$~10$^{-5}$.

Table~\ref{t_dmass} lists the global dust masses calculated
from the data in Tables~\ref{t_smm}, \ref{t_firsmmsed_1}, \ref{t_firsmmsed_n},
and \ref{t_firsmmsed_2}.   Dust masses calculated with steeper
emissivity laws are always higher than those calculated with shallower
emissivity laws for two reasons.  First, the fits to the spectral energy
distribution using the shallower emissivity function produce higher
temperatures than those fits using the steeper emissivity functions.  These
higher temperatures lead to higher values of B($\nu$,T), which then result in
lower calculated dust masses.   The second is the
difference in the dust grain emissivity values at 850~$\mu$m.  The
emissivity is higher for shallower emissivity functions, which leads to
lower calculated dust masses.  Essentially, a higher temperature or a
higher emissivity requires less dust to produce the measured submillimeter
flux.

Table~\ref{t_dmass} also presents global molecular gas masses calculated from
the data in \citet{yetal95} and the resulting global gas to dust mass ratios.
(We did examine gas to dust ratios for separate nuclear and disk regions,
but the comparison showed no clear distinction between nuclear and
disk ratios.)

\subsection{Discussion}

Inspection of Table~\ref{t_dmass} shows that the gas to dust ratios
ultimately depend on the emissivity function used to fit the data.
Perhaps even more importantly, when the emissivity coefficient is fixed
(i.e. proportional to $\lambda^{-1}$ or $\lambda^{-2}$), the range of gas to
dust ratios for these 8 galaxies varies by factors of 20 - 30.  With the
$\lambda^{-1}$ emissivities, the ratios range from 130 to 2700, and with the
$\lambda^{-2}$ emissivities and two temperature components, the ratios range
from 13 to 480.  However, when the emissivity coefficient is fit to the data,
the gas to dust ratios vary from 150 to 580, only a factor $\sim$~4.

We may check the plausibility of these calculated gas to dust ratios using the
Milky Way ratio as a benchmark.  The molecular gas to dust ratio in the Milky
Way has been determined to be 115 using COBE data \citep{sbbetal94}.  The
median gas to dust ratio for the $\lambda^{-1}$ emissivity fits is
approximately a factor of 6 higher than the ratio for the Milky Way, so the
single blackbody with the $\lambda^{-1}$ emissivity is unlikely to
accurately represent the physical conditions of the dust.  The
median ratio from the $\lambda^{-2}$ emissivity fit is within a factor of 3
of the Milky Way value, so these fits are more plausible. The median ratio
from the $\lambda^{-\beta}$ emissivity fits is also within a factor of 3, so
these fits are equally plausible.  \citet{dh02}, however, have demonstrated 
empirically that fitting one or two blackbodies to far-infrared and 
submillimeter data may lead to underestimates of the dust mass by as much 
as a factor of 10 when compared to more realistic models that fit a series of 
blackbodies to the data.  If we accept that this effect is present, 
the $\lambda^{-1}$ and
$\lambda^{-\beta}$ emissivity fits would produce gas to dust ratios near the
Milky Way's but the $\lambda^{-2}$ emissivity ratios would be much too low
to be plausible.

However, the range in ratios provides an additional criterion for comparing
the three fits.  The dust masses computed using both the
$\lambda^{-1}$ emissivity and the two-component model with very cold dust
give gas-to-dust ratios which vary from galaxy to galaxy by factors of 20
and 37 respectively.  In contrast, the range in the ratios for the
$\lambda^{-\beta}$ fits is less than a factor of 4.  Since these are all
nearby spiral galaxies, we would expect the basic properties of the
interstellar medium to be similar.  If the best description of the
dust is the one that gives the lowest dispersion in the gas-to-dust ratios,
and gas/dust ratios which are similar to that for the MIlky Way, then the
$\lambda^{-\beta}$ emissivity fits are clearly preferred.  The fits with the
fixed emissivities give implausible variations in the gas-to-dust ratios,
while the $\lambda^{-\beta}$ fits give a more acceptable range.

As a result, using both closeness to the Milky Way gas-to-dust ratio and the 
small dispersion of the ratios as an indication of the plausibility for the
gas-to-dust ratios, it is evident that the fits using the $\lambda^{-\beta}$
emissivity, i.e. the right-hand column of Table~\ref{t_dmass}, provide
results that are most likely to be correct for the galaxies in this sample.

We are now able to address the larger question of the lack of consensus in
the dust temperatures and even the number of different temperature
components in normal galaxies which was described in the Introduction.  It
is clear from the analysis above that one can fit spectral energy
distributions from 60 - 850 $\mu$m equally well by a) a single temperature
component with emissivity proportional to $\lambda^{-1}$, b) a single
temperature component with emissivity proportional to $\lambda^{-\beta}$, or
c) two temperature components, one of which is "cold," with a temperature
$\sim$~15~K, and the other with temperature $\sim$~30~K.  This range of
results is just what one finds in reviewing the literature on this subject
over the past 20 years.  We suggest it is the assumptions made about the
emissivity function in different studies which have largely produced the
wide range of conclusions about the dust temperatures and dust masses in
galaxies and the lack of consensus which has plagued this subject.  We have
shown that that the assumptions most commonly used lead to implausible
gas-to-dust ratios. The distinguishing feature of this study is that we have
tested the physical plausibility of various possible emissivity functions
rather than adhering rigidly to an assumed emissivity function and
calculating dust temperatures and other parameters.

So, having shown that the $\lambda^{-\beta}$ emissivity fits to the spectral
energy distributions imply more plausible dust masses and gas-to-dust ratios
than the other choices, what is the physical interpretation of this result?
First we note that in this model we use only one temperature component to
approximate the dust emission when there is almost certainly a range of dust
temperatures both within  star formation regions and in the diffuse
interstellar medium.  Does this make sense? A priori, a two-component model
might be more reasonable since a warmer component from star
formation regions and a  cooler component from dust heated by the diffuse
interstellar radiation  field are expected.  Since such models appear to
result in dust masses that are too large and gas-to-dust ratios that are
implausibly small, this suggests that the colder component in such models is
a less significant contributor than the two-component fits imply.  Since
only a single temperature gives plausible dust masses, we suggest that
if a second cold component to the dust emission from the diffuse
interstellar medium is present, it is relatively insignificant in terms of
both its contributions to the mass and  to the SED.   To distinguish its
contribution apparently requires better spectral coverage and photometric
precision in the submillimeter and millimeter regions than has been
available heretofore.

We must also consider the physical interpretation of a variable
emissivity law due to the varying $\beta$ we find for this galaxy sample.
The emissivity functions should be related to the average physical properties
of the dust, such as the composition and grain size distribution, and these
are expected to be relatively constant among nearby spiral galaxies.
Is it not inconsistent to argue that the emissivity functions can vary when we
also argue that the gas-to-dust ratios should be similar among spiral
galaxies?   We suggest two plausible interepretations for this result.
First, it may be that the grain properties do vary among these galaxies.
\citet{dh02} and references therein present data that demonstrate how
$\beta$ may depend on environment.  However, they argue that $\beta$ should
decrease as the temperature of dust increases, which is the opposite of the
trend we have found. Instead, we suggest that galaxies which have undergone
significant recent star formation activity and concomitant supernovae may
have a somewhat altered dust grain population. Supernova shock waves can
break down silicate grains with carbon mantles into individual silicate and
graphite grains. Silicate grains have different emissivities than the
graphite grains, so the average emissivity would change.  Furthermore, the
destruction of very large grains would the make the emissivity function
steeper. Breaking down the dust with supernova shock waves also suggests
some dust grain destruction \citep{dw97},  which is exacly what is seen in
these results for galaxies with steep emissivities.  An alternative
interpretation might be that the varying emissivity function is simply a
second fitting parameter, in addition  to a temperature, and this combination
better accounts for the range of dust temperatures and other dust properties
that are found in spiral galaxies.

\section{Conclusions}

We have shown that, for the 71 galaxies in the ISO Atlas sample, the 60 -
180~$\mu$m spectral energy distributions have mean temperatures of 31~K
for the $\lambda^{-1}$ emissitivities and 25~K for the $\lambda^{-2}$
emissitivities.  The blackbodies with $\lambda^{-1}$ emissitivities
generally fit the data better than the ones with
$\lambda^{-2}$ emissivities.  Analysis of the dust temperatures as a function
of morphological type reveals that spirals from Sa-Scd have remarkably
uniform dust temperatures which are virtually identical within the errors.
Dust temperatures are enhanced above the average for the spirals only in
E/SO galaxies and in galaxies with especially strong star formation
activity.  Data based on SEDs over this spectral range do not support the
existence of a substantial cold component at a temperature $<$~20~K
in a large fraction of spiral galaxies, in contrast to some earlier studies.

For the eight galaxies in the sample with the spectral energy distributions
measured from 60 - 850~$\mu$m, we showed that the 60 - 850~$\mu$m spectral
energy distributions could be fit equally well by: a) a single temperature
component with emissivity proportional to $\lambda^{-1}$, b) a single
temperature component with emissivity proportional to
$\lambda^{-\beta}$, or c) two temperature components, one of which is "cold,"
with a temperature $\sim$~15~K.  We suggest that the enormous
variation in reported dust temperatures and dust masses that characterize
the literature on spiral galaxies is largely a reflection of the variety of
assumptions for the emissivity functions employed.

To choose among these three assumptions we used the gas-to-dust ratio as a
criterion.  Only the temperatures derived using the single temperature
component with emissivity proportional to
$\lambda^{-\beta}$ produced gas-to-dust ratios within a rather narrow range
and one that was close to the gas-to-dust ratio for the Milky Way.  We
suggest that these results are most likely to represent the actual dust
properties of normal spiral galaxies.  In particular, we find the wide range
in inferred gas-to-dust ratios derived with the assumption of a "cold" dust
component less physically plausible than those ratios derived using a single
temperature component with temperatures derived using $\lambda^{-\beta}$,
with $\beta$ calculated for each galaxy.

\acknowledgements

G. J. B. would like to thank Iain Coulson and Loretta Dunne for their help with
the submillimeter observations and data processing; Daniel A. Dale for his
comments on this paper; and the staff of IRAS for providing the HIRES data.

The ISOPHOT data presented in this paper were reduced using PIA, which is a
joint development by the ESA Astrophysics Division and the ISOPHOT Consortium
with the collaboration of the Infrared Processing and Analysis Center (IPAC).
Contributing ISOPHOT Consortium institutes are DIAS, RAL, AIP, MPIK, and MPIA.

The JCMT is operated by the Joint Astronomy Center on behalf of the
UK Particle Physics and Astronomy Research Council, the Netherlands
Organization for Scientific Research, and the Canadian National Research
Council.

This research has been supported by NASA grants NAG 5-3370 and JPL 961566.

\clearpage
\onecolumn

\clearpage

\begin{deluxetable}{lcccccc}
\tablecolumns{7}
\tablewidth{0pc}
\tablecaption{Submillimeter Flux Density Measurements \label{t_smm}}
\tablehead{
\colhead{Galaxy} &
\multicolumn{2}{c}{450 $\mu$m Flux} &
\multicolumn{2}{c}{850 $\mu$m Flux} &
\colhead{log log D$_{25}$\tablenotemark{a}} &
\colhead{Data}
\\
\colhead{} &
\multicolumn{2}{c}{Density (Jy)} &
\multicolumn{2}{c}{Density (Jy)} \\
\colhead{} &
\colhead{45 arcsec} & \colhead{135 arcsec} &
\colhead{45 arcsec} & \colhead{135 arcsec} &
\colhead{} &
\colhead{Source}}
\startdata
NGC 4088 &	\nodata &	\nodata &
                 $0.26 \pm 0.03$ &     $0.98 \pm 0.10$ &
                 1.76 &     Observation \\
NGC 4414 &	$2.8 \pm 0.3$ &	      $7.3 \pm 0.7$ &
                 $0.36 \pm 0.04$ &     $0.84 \pm 0.08$ &
                 1.56 &     Archive \\
NGC 4631 &	$4.8 \pm 0.5$ &	      $18. \pm 0.2$ &
                 $0.54 \pm 0.05$  &    $1.89 \pm 0.19$ &
                 2.19 &     Archive \\
NGC 4826 &	\nodata &	\nodata &
                 $0.63 \pm 0.06$ &     $1.01 \pm 0.10$ &
                 2.00 &     Observation \\
NGC 5236 &	$5.5 \pm 0.5$ &       $9.7 \pm 0.9$ &
                 $0.82 \pm 0.08$ &     $1.36 \pm 0.14$ &
                 2.11 &     Archive \\
NGC 5713 &	\nodata &	\nodata &
                 $0.26 \pm 0.03$ &     $0.43 \pm 0.04$ &
                 1.44 &     Archive \\
NGC 5792 &	\nodata &	\nodata &
                 $0.18 \pm 0.02$ &     $0.33 \pm 0.03$ &
                 1.84 &     Archive \\
NGC 5907 &	\nodata &	\nodata &
                 $0.43 \pm 0.04$ &     $1.26 \pm 0.13$ &
                 2.10 &     Archive \\
\tablenotetext{a}{The optical diameter parameter defined in \citet{detal91}.}
\enddata
\end{deluxetable}

\clearpage

\begin{deluxetable}{lcc}
\tablecolumns{3}
\tablewidth{0pc}
\tablecaption{Temperatures from Far-Infrared Spectral Energy Distributions
\label{t_firsed}}
\tablehead{
\colhead{Galaxy} &
\multicolumn{2}{c}{Temperature (K)\tablenotemark{a}} \\
\colhead{} &
\colhead{$\lambda^{-1}$ emissivity} &
\colhead{$\lambda^{-2}$ emissivity}}
\startdata
NGC   55 &      26 &      22 \\
NGC  289 &      31 &      25 \\
NGC 1512 &      31 &      26 \\
NGC 3359 &      32 &      26 \\
NGC 3556 &      31 &      26 \\
NGC 3898 &      30 &      25 \\
NGC 4062 &      28 &      23 \\
NGC 4088 &      31 &      25 \\
NGC 4096 &      29 &      24 \\
NGC 4100 &      33 &      26 \\
NGC 4136 &      30 &      25 \\
NGC 4157 &      29 &      24 \\
NGC 4203 &      32 &      26 \\
NGC 4236 &      32 &      26 \\
NGC 4244 &      27 &      23 \\
NGC 4274 &      29 &      24 \\
NGC 4314 &      34 &      28 \\
NGC 4395 &      26 &      22 \\
NGC 4414 &      31 &      25 \\
NGC 4448 &      30 &      25 \\
NGC 4559 &      29 &      24 \\
NGC 4605 &      32 &      26 \\
NGC 4618 &      30 &      25 \\
NGC 4631 &      34 &      27 \\
NGC 4710 &      33 &      27 \\
NGC 4725 &      24 &      21 \\
NGC 4826 &      33 &      27 \\
NGC 4984 &      39 &      30 \\
NGC 5033 &      30 &      25 \\
NGC 5054 &      31 &      26 \\
NGC 5055 &      28 &      24 \\
NGC 5087 &      38 &      30 \\
NGC 5101 &      28 &      24 \\
NGC 5102 &      29 &      24 \\
NGC 5112 &      29 &      24 \\
NGC 5170 &      28 &      24 \\
NGC 5204 &      33 &      27 \\
NGC 5236 &      36 &      29 \\
NGC 5247 &      29 &      24 \\
NGC 5300 &      29 &      24 \\
NGC 5334 &      28 &      23 \\
NGC 5364 &      26 &      22 \\
NGC 5371 &      28 &      23 \\
NGC 5457 &      27 &      23 \\
NGC 5474 &      32 &      26 \\
NGC 5556 &      33 &      27 \\
NGC 5566 &      29 &      24 \\
NGC 5584 &      32 &      26 \\
NGC 5585 &      33 &      27 \\
NGC 5669 &      29 &      24 \\
NGC 5676 &      31 &      26 \\
NGC 5701 &      31 &      25 \\
NGC 5713 &      37 &      29 \\
NGC 5746 &      28 &      23 \\
NGC 5792 &      33 &      27 \\
NGC 5838 &      42 &      32 \\
NGC 5846 &      37 &      29 \\
NGC 5850 &      30 &      25 \\
NGC 5866 &      32 &      26 \\
NGC 5907 &      26 &      22 \\
NGC 5985 &      28 &      23 \\
NGC 6015 &      29 &      24 \\
NGC 6215 &      35 &      28 \\
NGC 6217 &      35 &      28 \\
NGC 6221 &      34 &      27 \\
NGC 6300 &      31 &      26 \\
NGC 6340 &      38 &      30 \\
NGC 6503 &      30 &      25 \\
NGC 6643 &      31 &      26 \\
NGC 6744 &      25 &      21 \\
NGC 6753 &      32 &      26 \\
\enddata
\tablenotetext{a}{Errors on the temperature measurements are 2~K.}
\end{deluxetable}

\clearpage

\begin{deluxetable}{lccc}
\tablecolumns{4}
\tablewidth{0pc}
\tablecaption{Dependence of Far-Infrared Dust Temperatures on
Morphological Type \label{t_firsedmorph}}
\tablehead{
\colhead{Morphological} & \colhead{Number} &
\colhead{Mean Temperatures (K)} & \colhead{Mean Temperatures (K)} \\
\colhead{Type} & \colhead{} &
\colhead{with $\lambda^{-1}$ Emissivity} &
\colhead{with $\lambda^{-2}$ Emissivity}}
\startdata
All &		71 &	30.9 $\pm$ 0.4 &	25.3 $\pm$ 0.3 \\
E - S0/a &	11 &	34.5 $\pm$ 1.4 &	27.5 $\pm$ 0.8 \\
Sa - Sab &	8 &	30.0 $\pm$ 1.1 &	25.0 $\pm$ 0.8 \\
Sb - Sbc &	19 &	30.3 $\pm$ 0.7 &	24.9 $\pm$ 0.5 \\
Sc - Scd &	25 &	30.2 $\pm$ 0.5 &	24.9 $\pm$ 0.3 \\
Sd - Irr &	8 &	30.9 $\pm$ 1.1 &	25.4 $\pm$ 0.8 \\
\enddata
\end{deluxetable}

\clearpage

\begin{deluxetable}{lccccc}
\tablecolumns{6}
\tablewidth{0pc}
\tablecaption{Results of Deconvolution Analysis on 180~$\mu$m Data
\label{t_180}}
\tablehead{
\colhead{} & \colhead{Measured 180~$\mu$m} &
\multicolumn{4}{c}{Deconvolved 180~$\mu$m Flux Densities (Jy)} \\
\colhead{Galaxy} & \colhead{Flux Densities} &
\multicolumn{2}{c}{in Components} &
\multicolumn{2}{c}{in Set Apertures} \\
\colhead{} & \colhead{(180$\arcsec$) (Jy)} &
\colhead{Central} & \colhead{Extended (135$\arcsec$)} &
\colhead{(45$\arcsec$)} & \colhead{(135$\arcsec$)}}
\startdata
NGC 4088 &     46.2 &          8.4 &     23. &           11. &     31.  \\
NGC 4414 &     54.5 &          22. &     23. &           25. &     45.  \\
NGC 4631 &     93.9 &          19. &     46. &           24. &     65.  \\
NGC 4826 &     51.8 &          18. &     34. &           34. &     50.  \\
NGC 5236 &     192. &          90. &     77. &           99. &     170. \\
NGC 5713 &     22.3 &          16. &     7.  &           17. &     23.  \\
NGC 5792 &     15.9 &          8.7 &     6.0 &           9.3 &     15.  \\
NGC 5907 &     44.1 &          9.1 &     22. &           12. &     31.  \\
\enddata
\end{deluxetable}

\clearpage

\begin{deluxetable}{lcccccc}
\tablecolumns{7}
\tablewidth{0pc}
\tablecaption{Temperatures with $\lambda^{-1}$ Emissivity
\label{t_firsmmsed_1}}
\tablehead{
\colhead{Galaxy} &
\multicolumn{2}{c}{Global} &
\multicolumn{2}{c}{Nuclear} &
\multicolumn{2}{c}{Disk}\\
\colhead{} &
\colhead{Temp. (K)\tablenotemark{a}} & \colhead{Reduced $\chi^2$} &
\colhead{Temp. (K)\tablenotemark{a}} & \colhead{Reduced $\chi^2$} &
\colhead{Temp. (K)\tablenotemark{a}} & \colhead{Reduced $\chi^2$}}
\startdata
NGC 4088 &	33. & 0.014 & 	35. & 0.44 & 	32. & 0.17 \\
NGC 4414 &	34. & 3.0 & 	35. & 0.95 & 	33. & 12. \\
NGC 4631 &	35. & 3.4 & 	37. & 3.5 &	34. & 3.7 \\
NGC 4826 &	35. & 2.0 & 	36. & 2.2 & 	32. & 2.5 \\
NGC 5236 &	45. & 10. & 	48. & 7.0 & 	41. & 18. \\
NGC 5713 &	39. & 1.6 & 	41. & 2.9 & 	34. & 0.64 \\
NGC 5792 &	35. & 0.79 & 	37. & 1.4 & 	30. & 1.26 \\
NGC 5907 &	27. & 0.33 & 	28. & 0.46 & 	27. & 0.26 \\
\enddata
\tablenotetext{a}{Errors on the temperature measurements are 2~K.}
\end{deluxetable}

\clearpage

\begin{deluxetable}{lccccccccc}
\tablecolumns{10}
\tablewidth{0pc}
\tablecaption{Temperatures with $\lambda^{-\beta}$ Emissivity
\label{t_firsmmsed_n}}
\tablehead{
\colhead{Galaxy} &
\multicolumn{3}{c}{Global} &
\multicolumn{3}{c}{Nuclear} &
\multicolumn{3}{c}{Disk}\\
\colhead{} &
\colhead{Temp.} & \colhead{$\beta$} &
\colhead{Reduced} &
\colhead{Temp.} & \colhead{$\beta$} &
\colhead{Reduced} &
\colhead{Temp.} & \colhead{$\beta$} &
\colhead{Reduced}\\
\colhead{} &
\colhead{(K)\tablenotemark{a}} & \colhead{} & \colhead{$\chi^2$} &
\colhead{(K)\tablenotemark{a}} & \colhead{} & \colhead{$\chi^2$} &
\colhead{(K)\tablenotemark{a}} & \colhead{} & \colhead{$\chi^2$}}
\startdata
NGC 4088 &	34. & 1.0 & 0.0055 & 	33. & 1.1 & 0.11 &
	34. & 0.9 & 0.0030 \\
NGC 4414 &	29. & 1.4 & 0.52 & 	31. & 1.3 & 0.66 &
	26. & 1.7 & 5.4 \\
NGC 4631 &	35. & 1.0 & 5.1 &  	34. & 1.2 & 3.8 &
	35. & 0.9 & 5.9 \\
NGC 4826 &	31. & 1.3 & 0.011 & 	32. & 1.3 & 0.11 &
	28. & 1.4 & 0.20 \\
NGC 5236 &	30. & 1.9 & 0.19 &	33. & 1.7 & 0.80 &
	26. & 2.1 & 0.69 \\
NGC 5713 &	34. & 1.3 & 0.022 &	35. & 1.4 & 0.15 &
	31. & 1.2 & 0.32 \\
NGC 5792 &	32. & 1.2 & 0.28 & 	34. & 1.2 & 0.82 &
	27. & 1.3 & 0.17 \\
NGC 5907 &	28. & 0.9 & 0.45 &	29. & 0.9 & 0.83 &
	28. & 0.9 & 0.28 \\
\enddata
\tablenotetext{a}{Errors on the temperature measurements are 2~K.}
\end{deluxetable}

\clearpage

\begin{deluxetable}{lcccccc}
\tablecolumns{7}
\tablewidth{0pc}
\tablecaption{Temperatures with $\lambda^{-2}$ Emissivity
\label{t_firsmmsed_2}}
\tablehead{
\colhead{Galaxy} &
\multicolumn{2}{c}{Global} &
\multicolumn{2}{c}{Nuclear} &
\multicolumn{2}{c}{Disk}\\
\colhead{} &
\colhead{Temp. (K)\tablenotemark{a}} & \colhead{Reduced $\chi^2$} &
\colhead{Temp. (K)\tablenotemark{a}} & \colhead{Reduced $\chi^2$} &
\colhead{Temp. (K)\tablenotemark{a}} & \colhead{Reduced $\chi^2$}}
\startdata
NGC 4088 &	13. \& 29. & \nodata &		14. \& 30. & \nodata &
			12. \& 28. & \nodata  \\
NGC 4414 &	15. \& 28. & 0.031 & 		7. \& 27. & 0.70 &
			18. \& 28. & 6.1 \\
NGC 4631 &	12. \& 30. & 2.5 & 		14. \& 31. & 1.9 &
			12. \& 30. & 2.9 \\
NGC 4826 &	14. \& 28. & \nodata &		15. \& 30. & \nodata &
			5. \& 25. & \nodata  \\
NGC 5236 &	29. & 0.29 & 			30. & 1.6 &
			27. & 0.78      \\
NGC 5713 &	14. \& 31. & \nodata &		16. \& 33. & \nodata &
			6. \& 27. & \nodata  \\
NGC 5792 &	15. \& 30. & \nodata &		17. \& 34. & \nodata &
			9. \& 24. & \nodata  \\
NGC 5907 &	13. \& 26. & \nodata &		14. \& 27. & \nodata &
			13. \& 25. & \nodata  \\
\enddata
\tablenotetext{a}{Errors on the temperature measurements are 2~K.}
\end{deluxetable}

\clearpage

\begin{deluxetable}{lccccccc}
\tablecolumns{8}
\tablewidth{0pc}
\tablecaption{Global Dust Masses and Gas to Dust Ratios \label{t_dmass}}
\tablehead{
\colhead{Galaxy} &
\multicolumn{3}{c}{Dust Masses (M$_\odot$)} &
\colhead{Molecular Gas} & \multicolumn{3}{c}{Gas / Dust Ratios}\\
\colhead{} &
\colhead{$\lambda^{-1}$} &
\colhead{$\lambda^{-2}$} &
\colhead{$\lambda^{-\beta}$} &
\colhead{Masses(M$_\odot$)} &
\colhead{$\lambda^{-1}$} &
\colhead{$\lambda^{-2}$} &
\colhead{$\lambda^{-\beta}$} \\
\colhead{} & \colhead{emissivity} &
\colhead{emissivity} & \colhead{emissivity} &
\colhead{} & \colhead{emissivity} &
\colhead{emissivity} & \colhead{emissivity}}
\startdata
NGC 4088 &
	4.8 $\times$ 10$^6$ & 	5.7 $\times$ 10$^7$ &   4.6 $\times$ 10$^6$ &
	1.5 $\times$ 10$^9$ &	310. &	 26. &  330.  \\
NGC 4414 &
	1.3 $\times$ 10$^6$ & 	1.2 $\times$ 10$^7$ &   2.6 $\times$ 10$^6$ &
	1.5 $\times$ 10$^9$ &	1200. &	130. &  580.  \\
NGC 4631 &
	1.4 $\times$ 10$^6$ & 	2.1 $\times$ 10$^7$ &   1.4 $\times$ 10$^6$ &
	4.4 $\times$ 10$^8$ &	310. &	 21. &  310.  \\
NGC 4826 &
	2.7 $\times$ 10$^5$ & 	2.7 $\times$ 10$^6$ &   4.5 $\times$ 10$^5$ &
	1.8 $\times$ 10$^8$ &	670. &	 67. &  400.  \\
NGC 5236 &
	3.5 $\times$ 10$^5$ & 	2.0 $\times$ 10$^6$ &   1.7 $\times$ 10$^6$ &
	9.5 $\times$ 10$^8$ &	2700. &	480. &  560.  \\
NGC 5713 &
	5.5 $\times$ 10$^6$ & 	6.5 $\times$ 10$^7$ &   9.3 $\times$ 10$^6$ &
	3.0 $\times$ 10$^9$ &	550. &	 46. &  320.  \\
NGC 5907 &
	6.2 $\times$ 10$^6$ & 	5.9 $\times$ 10$^7$ &   5.2 $\times$ 10$^6$ &
	7.8 $\times$ 10$^8$ &	130. &	 13. &  150.  \\
\enddata
\end{deluxetable}

\end{document}